# I CAN HAS SUPERCOMPUTER?

A Novel Approach to Teaching Parallel and Distributed Computing Concepts Using a Meme-Based Programming Language

David A. Richie
Brown Deer Technology
Forest Hill, MD
drichie@browndeertechnology.com

James A. Ross
U.S. Army Research Laboratory
Aberdeen Proving Ground, MD
james.a.ross176.civ@mail.mil

*Abstract*—A novel approach is presented to teach the parallel and distributed computing concepts of synchronization and remote memory access. The single program multiple data (SPMD) partitioned global address space (PGAS) model presented in this paper uses a procedural programming language appealing to undergraduate students. We propose that the amusing nature of the approach may engender creativity and interest using these concepts later in more sober environments. Specifically, we implement parallel extensions to LOLCODE within a source-to-source compiler sufficient for the development of parallel and distributed algorithms normally implemented using conventional high-performance computing languages and APIs.

## I. INTRODUCTION

The modern undergraduate demographic has been born into an internet culture where poking fun at otherwise serious issues is considered cool. Internet memes are the cultural currency by which ideas are transmitted through younger audiences. This reductionist approach using humor is very effective at simplifying often complex ideas. Internet memes have a tendency to rise and fall in cycles, and as with most things placed on the public internet, they never really go away. In 2007, the general-purpose programming language LOLCODE was developed and resembled the language used in the LOLCAT meme which includes photos of cats with comical captions [1], and with deliberate pattern-driven misspellings and common abbreviations found in texting and instant messenger communications.

The nature of the LOLCODE language is perhaps best understood by, and most notably identified with, its replacement of the standard C directive `#include <stdio.h>` with the statement `CAN HAS STDIO?` The minimalist, themed language syntax makes the code read like a joke, making it approachable for a younger generation. For some time, a community developed around the language and many tools were built, including a LOLCODE interpreter [2]. However, the language has not evolved much since its creation. To the best of our knowledge, no esoteric or meme-based programming languages support parallelism or distributed computing concepts.

Computationally-demanding problems that either cannot be solved on a single machine or cannot execute fast enough must use parallel and distributed programming, however, it can be challenging. Parallel and distributed language development remains an open problem today and is often complicated by application programming interfaces (APIs) or compiler directives rather than the programming language itself. Relatively rapid changes in hardware often come with new methods for programming them so knowledge gained with one platform is not directly transferable to another. There are basic concepts commonly found in most parallel programming models and we recognize that concepts are more important than the implementation specifics. We propose minor language syntax extensions to the LOLCODE language in order to teach these concepts. These extensions enable parallel synchronization and remote memory access within a single program multiple data (SPMD) partitioned global address space (PGAS) model to be explored as a toy in this reductionist manner. Unlike humorous efforts to bring parallelism to the language in the past [3], this effort is not simply creating bindings to the parallel message passing interface (MPI). Instead we propose language extensions that create parallel programming semantics within the spirit of the original LOLCODE specification. The extensions enable software to scale from inexpensive low-power parallel education platforms to the largest supercomputers.

## II. SOFTWARE IMPLEMENTATION AND HARDWARE

We have developed a LOLCODE compiler, including our proposed language extensions, enabling the compilation of parallel and distributed LOLCODE applications on virtually any platform with a C compiler and OpenSHMEM library [4]. We have chosen to initially target the Parallella board [5] since it is an inexpensive educational or developmental platform with a number of hardware technologies that may be explored and has been used in a number of undergraduate computer science courses as an introduction to parallel programming and platforms [6]. Hardware features found on a Parallella board include a dual-core ARM processor, a Xilinx FPGA, GPIO pins, and a 16-core Adapteva Epiphany-III coprocessor [7]. The Epiphany-III is a low-power 2D RISC array architecture with a network on chip (NoC) and may be thought of, and programmed, as a cluster on a chip. We have previously developed the COPRTHR-2 SDK [8] and the OpenSHMEM implementation [9] to support the Parallella platform and our parallel LOLCODE implementation leverages this software.

The LOLCODE compiler is a source-to-source compiler, written in C, and translates LOLCODE with parallel extensions to C with OpenSHMEM routines to implement the parallel and distributed computing extensions. After this translation, a standard C compiler is used to compile the code into an executable program. We demonstrate parallel LOLCODE applications running on the $99 Parallella board, with the 16-core Adapteva Epiphany coprocessor, as well as (a portion of) the $30 million US Army Research Laboratory's 101,312-core Cray XC40 production supercomputer.

We assert that the concepts of text-based code development and the parallel methods explored here represent the foundational knowledge and understanding needed for software development on high performance computing (HPC) systems. The abstract underpinnings of the computer science concepts of parallel concurrency, synchronization, and memory movement are some of the most difficult to understand and appreciate. Therefore, teaching these concepts to students is of a sufficient challenge and importance to warrant innovative approaches. By contrast, the particulars of programming languages and parallel programming APIs are of secondary importance for students interested in high-performance computing and may be easier to understand intuitively once the foundational concepts are learned.

### A. Memory Model

The underlying memory model used for this work is based on PGAS and the execution model is based on SPMD. Each parallel processing element (PE) runs a single program with symmetric memory allocation (symmetric shared arrays and statically declared variables) as shown in Figure 1.

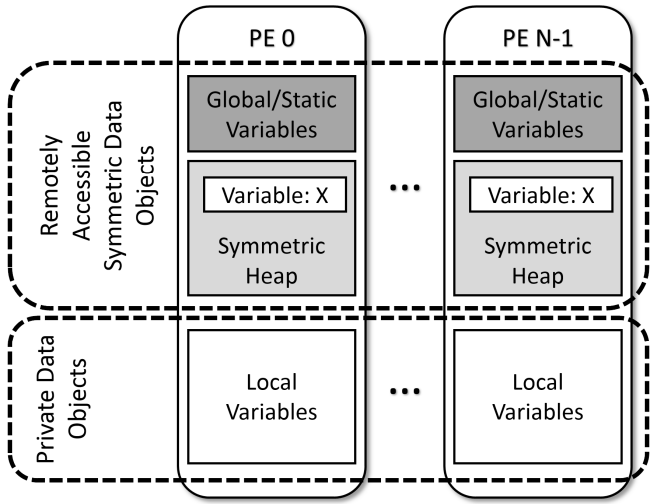

Figure 1: The PGAS memory model used in this work

The PGAS programming model with symmetric memory across a system may be one of the simplest models for distributed computing. The one-sided message passing interface, OpenSHMEM, exposes distributed memory resources. The U.S. Army Research Laboratory (ARL) has recently published an OpenSHMEM implementation for Epiphany which enabled a standard parallel programming interface to the 16-core Adapteva Epiphany coprocessor included with the Parallella board. ARL views the Epiphany architecture as a serious contender for future HPC and embedded computing efforts due to its efficiency and scalability. By leveraging the standard OpenSHMEM interface, portable yet efficient applications may be developed. It is a result of our recent efforts in this area, and the simplicity of the programming model, that we believe it is reasonable to use OpenSHMEM as a backend for the parallel LOLCODE extensions.

Although the OpenSHMEM API may be powerful, our extensions focus on a minimal subset for the enumeration of process elements, parallel synchronization, and memory management. Other OpenSHMEM routines are used implicitly in the backend but do not have a direct language analog in order to simplify the programming model.

### B. Technical Challenges

The first technical challenge encountered in this work involved identifying extensions needed to support simple examples of parallel code commonly used for demonstrating computational algorithms. This work is based on the LOLCODE-1.2 specification as implemented in the LOLCODE project interpreter [2]. The language uses dynamic typing which we extend to support statically typed variables as a transition to a compiled, rather than interpreted language. We also make minor extensions that do not alter the spirit of the language such as allowing multiple clauses in declarations and adding built-in keywords for random numbers and additional math functions. A significant extension is the introduction of support for real arrays that can be dynamically sized and indexed with clean syntax. Support for arrays was notably lacking in the language and supported weakly with rather awkward and unintuitive syntax. The lack of support for arrays was thoroughly discussed, yet arrays are an important programming concept and must be available in any language used for teaching parallel algorithm concepts.

The second technical challenge was defining language extensions to support parallel programming. The key concepts that must be supported are parallel concurrency, synchronization, and distributed shared data access. Simply introducing hooks or bindings for OpenSHMEM or other parallel APIs would miss the point and the opportunity that the LOLCODE language presents as a compelling teaching tool. Therefore a simple yet reasonably efficient parallel programming model was developed that is consistent with the spirit of the language.

The third technical challenge was implementing the language extensions so that they could be used for targeting platforms like Parallella. In order to accomplish this, a source-to-source LOLCODE compiler was developed using lex [10] and yacc [11], with support for the proposed language extensions. The source-to-source compiler is portable and will support architectures with an available C compiler and OpenSHMEM implementation. Using a compiler for LOLCODE is more flexible and efficient than an interpreter and is better suited for use on parallel computing platforms.

### III. PRIMER ON LOLCODE

LOLCODE may be misclassified as an esoteric language since its structure, semantics, and syntax are simple and easy to understand despite the somewhat humorous syntax. As designed it is a simple dynamically typed interpreted language easily amenable to compilation. Simple interpreters have been developed to parse the language that may be easily understood by undergraduate computer science students [2]. The main keywords of the language are capitalized and often deliberately misspelled. It supports integer (`NUMBR`), floating point (`NUMBAR`), string (`YARN`), and Boolean (`TROOF`) types. Arrays are only supported as an unofficial extension to structs and use a somewhat awkward syntax, which we address in this work. Additionally, LOLCODE supports a limited set of logical, arithmetic, and string operations. Variables are declared and their types and values may be explicitly set or updated at any time. Control flow is supported with the standard programing constructs of conditionals, loops and case statements. The equivalent of functions are supported for modular programming. Table 1 presents the basic syntax for the LOLCODE language.

TABLE I. BASIC SYNTAX FOR LOLCODE LANGUAGE

| | |
|---|---|
| `HAI [version]` | Begins a program |
| `KTHXBYE` | Terminates program |
| `BTW Comment` | Single line comment |
| `OBTW`<br>`[Comment]`<br>`TLDR` | Multi line comment |
| `CAN HAS [library]?` | Includes the `STDIO`, `STRING`, `SOCKS`, and `STDLIB` libraries |
| `VISIBLE [arg]` | Prints `arg` to standard output |
| `GIMMEH [var]` | Reads `var` from standard input |
| `I HAS A [var]` | Declare variable |
| `I HAS A [var] ITZ [value]` | Declare and initialize variable |
| `I HAS A [var] ITZ A [type]` | Declare typed variable |
| `[var] R [value]` | Assigns value to variable |
| `[OP] [expression] AN [expression]`<br><br>where `OP` is one of:<br>`BOTH SAEM, DIFFRINT, BIGGER, SMALLR, SUM OF, PRODUKT OF, QUOSHUNT OF, or MOD OF` | Performs logical and arithmetic operations *equal*, *not equal*, *greater than*, *less than*, *add*, *subtract*, *multiply*, *divide*, or *modulo*, respectively. |
| `MAEK [expression] A [type]` | Explicitly casts `expression` to `type` |
| `[variable] IS NOW A [type]` | Explicitly casts `variable` to `type` |
| `SRS [string]` | Interprets `string` as identifier |
| `[expression], O RLY?`<br>`  YA RLY`<br>`    [code]`<br>`  NO WAI`<br>`    [code]`<br>`OIC` | If/else statement block |
| `[expression], WTF?`<br>`  OMG [value]`<br>`    [code]`<br>`  OMG [value]`<br>`    [code]`<br>`  OMGWTF`<br>`    [code]`<br>`OIC` | Switch statement to compare the expression with literal values. A `GTFO` statement acts as a break, otherwise the next `OMG` block is executed. The `OMGWTF` block is the default. |
| `IM IN YR [label]`<br>`(UPPIN|NERFIN YR [var])`<br>`(TIL|WILE [expression]))`<br>`  [code]`<br>`IM OUTTA YR [label]` | A loop construct to increase (`UPPIN`) or decrease (`NERFIN`) a variable until (`TIL`) or while (`WILE`) an expression is true. `GTFO` breaks the loop. |
| `...` | The triple dot enables continuation from a single line |
| `[statement],[statement]` | Multiple statements may be separated by a comma |

### IV. EXTENSIONS FOR DISTRIBUTED PARALLELISM

Perhaps more than other programming languages, there is special importance to the form and syntax of any proposed language extension since it is not sufficient to simply enable functionality. The extensions must conform to the underlying internet meme and instant messaging slang upon which the language is based. We introduce several new semantic concepts which perform symmetric operations for distributed parallelism within the context of a SPMD PGAS programming model. These concepts are implemented by a limited set of OpenSHMEM routines on the backend. Within this programming model, a processing element, or PE, is used to define a parallel thread within a symmetric execution environment. Table 2 lists the modifications for parallel synchronization and remote data access. Symmetric shared arrays are defined along with their local and remote addressing methods.

TABLE II. PARALLEL AND DISTRIBUTED COMPUTING EXTENSIONS

| | |
|---|---|
| `MAH FRENZ` | Integer value for the total number of parallel PEs running |
| `ME` | Integer value identifying the PE of the executing thread |
| `IM SRSLY MESIN WIF [var]` | Acquire the global exclusive lock associated with `var` |
| `IM MESIN WIF [var], O RLY?`<br>`  YA RLY,`<br>`    [code]`<br>`OIC` | Test the lock associated with `var`, non-blocking |
| `DUN MESIN WIF [var]` | Release the lock associated with `var` |
| `HUGZ` | Collective barrier operation |
| `TXT MAH BFF [expr], [statement]` | Predicate statement such that remote references target the address-space of the PE `expr` |
| `TXT MAH BFF [expr] AN STUFF`<br>  `[code]`<br>`TTYL` | Predicate all statements in `code` such that remote references target the address-space of the PE `expr`. |
| `I HAS A [var] ITZ SRSLY A [type]` | Declare a variable statically typed to `type` |
| `WE HAS A [var] ITZ SRSLY A [type] AN IM SHARIN IT` | Declare a symmetric shared `var` statically typed to `type` |
| `WE HAS A [var] ITZ SRSLY LOTZ A [type]S AN THAR IZ [size]` | Declare a symmetric shared array `var` statically typed to `type` with `size` elements |
| `UR\|MAH [var]` | Evaluate `var` within the remote (`UR`) or local (`MAH`) address-space, only valid within a statement that is predicated |
| `[var]'Z [expr]` | Evaluate to the array element of `var` at the index `expr` |

In order to improve programmability and support for typical computational algorithms, additional extensions are introduced. They were effectively used in a canonical parallel distributed n-body calculation.

TABLE III.     ADDITIONAL LOLCODE EXTENSIONS

| | |
|---|---|
| `WHATEVR` | Random integer, `rand()` |
| `WHATEVAR` | Random floating point, `randf()` |
| `SQUAR OF [var]` | Power of 2, `var * var` |
| `UNSQUAR OF [var]` | Square root operation, `sqrt(var)` |
| `FLIP OF [var]` | Reciprocal operation (`1/var`) |

## V. PARALLEL PROGRAMING MODEL FOR LOLCODE

The parallel programing model which the extensions support is based on SPMD execution in which the same program is executed in parallel with multiple processing elements (PEs). Within this model, threads of concurrent execution are enumerated and each thread must be able to determine the total number of parallel threads being executed across the PEs as well as its own identity within the enumerated set. Within our parallel extensions to LOLCODE we refer to threads as *friends*. The expression `MAH FRENZ` evaluates to an integer representing the total number of parallel threads of execution. The identity of the executing thread is simply represented by the expression `ME`, which evaluates to an integer representing the unique identity within the enumeration.

Symmetric memory allocation with shared access consistent with a PGAS model is supported for variables and arrays allocated within each thread. This symmetric shared memory model is introduced with the scope identifier `WE` analogous to the existing LOLCODE scope identifier `I`. For example, the statement `WE HAS A x ITZ SRSLY A NUMBR` declares the symmetric variable `x` such that it may be accessed from any thread in a manner consistent with the backend SHMEM implementation. Symmetric shared arrays can be similarly declared with a statement such as `WE HAS A x ITZ SRSLY LOTZ A NUMBRS AN THAR IZ 100`, which declares an array of 100 elements symmetrically for each PE.

Synchronization between threads of parallel execution is a key concept in parallel and distributed computing models. One of the most common operations used to support synchronization is a barrier that must be entered by all threads before any thread can continue. Barriers are commonly used to guarantee that new data has been written to or received from a remote PE. We introduce the keyword `HUGZ` as a parallel barrier for LOLCODE, such that all threads (`MAH FRENZ`) must enter the barrier (`HUGZ`) before any can proceed.

Another basic synchronization concept involves mutually exclusive locks (mutexes) that must be acquired before a particular shared resource can be safely used or modified. We do not introduce explicit mutexes or locks into the language but instead use implied locks on variables and arrays that are declared as a shared resource. For example, the statement `WE HAS A x ITS A NUMBR AN IM SHARIN IT` will declare a symmetric variable `x` of integer type and also declare a hidden lock to protect the resource that may be acquired and released by association. If a thread will alter the `x` value it must first implicitly acquire the lock on it, and then subsequently release the lock when the critical operation is completed. The statements `IM SRLSY MESIN WIF x` and `DUN MESIN WIF x` are introduced as extensions to implicitly acquire and release a lock protecting the shared resource `x`. A non-blocking trylock operation is also available with the statement `IM MESIN WIF, O RLY?`

For example, in the code fragment below, a non-blocking attempt to acquire a lock (trylock) protecting the variable `x` is first performed, and if the attempt fails, a blocking attempt (lock) is made, before variable `x` is modified, and finally the lock is released (unlock),

```
IM SRSLY MESIN WIF X, O RLY?
NO WAI,
  IM MESIN WIF X
OIC
X R new_value
DUN MESIN WIF X
```

We introduce thread predication as a simple model for disambiguating symmetrically allocated shared data between threads to support a PGAS model for accessing remote data. Any statement immediately preceded by the statement `TXT MAH BFF [expr]` will be thread-predicated such that all references to remote variables or arrays will refer to the address-space of the thread identified by `expr`. Additionally we introduce the qualifier `UR` that may be applied to any variable or array such that the reference is to the remote address-space. Use of the corresponding qualifier `MAH` for references to the threads local variable or array is optional but makes thread-predicated statements more clear and logical. As an example, the statement `TXT MAH BFF k, MAH x R UR x` will cause the local variable `x` to be assigned the remote value of `x` in the address-space of thread `k`.

This semantic is more powerful than a straightforward data movement directive or explicit message passing since complex statements can be constructed with multiple remote references. For example, in the following statement, variable `x` is assigned the sum of remote variables `y` and `z`.

```
TXT MAH BFF k, MAH x R SUM OF UR y AN UR z
```

Additionally, an entire code-block can be thread predicated with the construct,

```
TXT MAH BFF k AN STUFF
  [code-block]
TTYL
```

All statements within the `code-block` will be thread-predicated so remote references refer to thread `k` (or PE `k`).

## VI. PARALLEL AND DISTRIBUTED MEMORY EXAMPLES

Demonstrating collective synchronization, point-to-point synchronization, and distributed memory methods as examples is essential for demonstrating the concepts. There are sufficient examples for the standard language available on the web so they are avoided in the following sections.

### A. Initialization and Symmetric Memory Allocation

OpenSHMEM requires initialization, but because every parallel LOLCODE program will call it, the initialization occurs transparently at the start of the program. Thus, symmetric allocation begins immediately within a program.

The following code fragment demonstrates identifying the processing element (`pe`) of the executing thread and the total number of processing elements (`n_pes`), allocating symmetric shared memory (`array[32]`), calculating the next processing element (`next_pe`) for a circular message transfer, and then copying the data of the neighboring `array` into the local `array` with the assignment operator.

```
I HAS A pe ITZ A NUMBR AN ITZ ME
I HAS A n_pes ITZ A NUMBR AN ITZ MAH FRENZ
WE HAS A array ITZ SRSLY LOTZ A NUMBRS ...
  AN THAR IZ 32
I HAS A next_pe ITZ A NUMBR ...
  AN ITZ SUM OF pe AN 1
next_pe R MOD OF next_pe AN n_pes
TXT MAH BFF next_pe, MAH array R UR array
```

### B. Parallel Syncrhonization with Locks

Protecting data in a multi-threaded programming model requires locks. The phrase `IM SHARIN IT` creates an implicit lock for the associated shared data so that the latter can be protected. The code fragment below demonstrates the concepts of allocating a symmetric shared data, protecting the data with a lock, modifying the remote value, and releasing the lock.

```
WE HAS A x ITZ A NUMBR AN IM SHARIN IT
TXT MAH BFF k AN STUFF
  IM MESIN WIF UR x
  x R SUM OF x AN 1
  DUN MESIN WIF UR x
TTYL
```

### C. Barriers and Message Passing

Barriers are collective operations that synchronize all PEs to meet at the same point within the program. The example below demonstrates collective data movement with synchronization between operations to prevent non-deterministic behavior.

```
TXT MAH BFF k, UR b R MAH a
HUGZ
c R SUM OF a AN b
```

Figure 2 represents this symmetric parallel operation. Each PE copies the local value of `a` to the variable `b` of the remote PE `k`. Subsequently, each PE calculates the local sum of `a` and `b` storing the result in `c`. The local PE memory is being concurrently accessed by other remote PEs as depicted by the incomplete arrows on the left and the right sides of Figure 2. Without synchronization, the program cannot prevent fast PEs from calculating the sum before their `b` value has been updated by the remote PE. This type of synchronization issue is typical for distributed memory applications found on HPC systems.

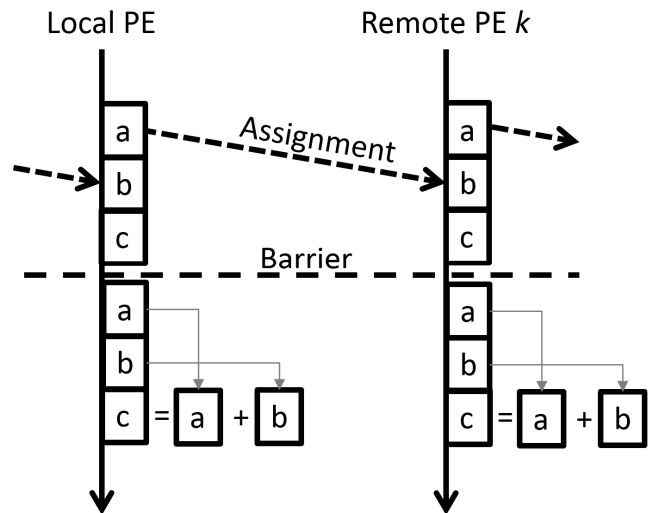

Figure 2: Visualization of symmetric parallel data movement

### D. Full Listing for Parallel 2D N-body Application

As a demonstration of a canonical HPC application, we implemented a parallel n-body program in LOLCODE, reduced to two dimensions here for simplicity. The source listing follows and demonstrates parallel symmetric memory allocation, initialization, remote memory access, and parallel synchronization with barriers.

```
HAI 1.2
OBTW
* 2D N-Body algorithm: propagate particles
* subject to Newtonian dynamics written in
* LOLCODE with parallel and other extensions.
TLDR

I HAS A little_time ITZ SRSLY A NUMBAR ...
   AN ITZ 0.001

I HAS A x ITZ SRSLY A NUMBAR
I HAS A y ITZ SRSLY A NUMBAR
I HAS A vx ITZ SRSLY A NUMBAR
I HAS A vy ITZ SRSLY A NUMBAR
I HAS A ax ITZ SRSLY A NUMBAR
I HAS A ay ITZ SRSLY A NUMBAR
I HAS A dx ITZ SRSLY A NUMBAR
I HAS A dy ITZ SRSLY A NUMBAR
I HAS A inv_d ITZ SRSLY A NUMBAR
I HAS A f ITZ SRSLY A NUMBAR

I HAS A vel_x ITZ SRSLY LOTZ A NUMBARS ...
   AN THAR IZ 32
I HAS A vel_y ITZ SRSLY LOTZ A NUMBARS ...
   AN THAR IZ 32
I HAS A tmppos_x ITZ SRSLY LOTZ A NUMBARS ...
   AN THAR IZ 32
I HAS A tmppos_y ITZ SRSLY LOTZ A NUMBARS ...
   AN THAR IZ 32

WE HAS A pos_x ITZ SRSLY LOTZ A NUMBARS ...
   AN THAR IZ 32 AN IM SHARIN IT
WE HAS A pos_y ITZ SRSLY LOTZ A NUMBARS ...
   AN THAR IZ 32 AN IM SHARIN IT

VISIBLE "HAI ITZ " ME " I HAS PARTICLZ 2 MUV"

HUGZ

IM IN YR loop UPPIN YR i TIL BOTH SAEM i AN 32
   pos_x'Z i R SUM OF ME AN WHATEVAR
   pos_y'Z i R SUM OF ME AN WHATEVAR
   vel_x'Z i R QUOSHUNT OF SUM OF ME ...
      AN WHATEVAR AN 1000
   vel_y'Z i R QUOSHUNT OF SUM OF ME ...
      AN WHATEVAR AN 1000
IM OUTTA YR loop

IM IN YR loop UPPIN YR time TIL BOTH SAEM ...
   time AN 10

   IM IN YR loop UPPIN YR i TIL BOTH SAEM ...
      i AN 32
      x R pos_x'Z i
      y R pos_y'Z i
      vx R vel_x'Z i
      vy R vel_y'Z i
      ax R 0
      ay R 0
      IM IN YR loop UPPIN YR j TIL ...
         BOTH SAEM j AN 32
         DIFFRINT i AN j, O RLY?
         YA RLY,
            dx R DIFF OF pos_x'Z i AN pos_x'Z j
            dy R DIFF OF pos_y'Z i AN pos_y'Z j
            dx R PRODUKT OF dx AN dx
            dy R PRODUKT OF dy AN dy
            inv_d R FLIP OF UNSQUAR OF ...
               SUM OF dx AN dy
            f R PRODUKT OF inv_d AN ...
               SQUAR OF inv_d
            ax R SUM OF ax AN PRODUKT OF dx AN f
            ay R SUM OF ay AN PRODUKT OF dy AN f
         OIC
      IM OUTTA YR loop

      IM IN YR loop UPPIN YR k TIL ...
         BOTH SAEM k AN MAH FRENZ
         DIFFRINT k AN ME, O RLY?
         YA RLY,
            IM IN YR loop UPPIN YR j TIL ...
               BOTH SAEM j AN 32
               TXT MAH BFF k AN STUFF,
                  dx R DIFF OF pos_x'Z i AN ...
                     UR pos_x'Z j
                  dy R DIFF OF pos_y'Z i AN ...
                     UR pos_y'Z j
               TTYL
               dx R PRODUKT OF dx AN dx
               dy R PRODUKT OF dy AN dy
               inv_d R FLIP OF UNSQUAR OF ...
                  SUM OF dx AN dy
               f R PRODUKT OF inv_d AN ...
                  SQUAR OF inv_d
               ax R SUM OF ax AN PRODUKT OF ...
                  dx AN f
               ay R SUM OF ay AN PRODUKT OF ...
                  dy AN f
            IM OUTTA YR loop
         OIC
      IM OUTTA YR loop

      x R SUM OF x AN SUM OF PRODUKT OF vx ...
         AN little_time AN PRODUKT OF 0.5 ...
         AN PRODUKT OF ax AN SQUAR OF ...
         little_time
      y R SUM OF y AN SUM OF PRODUKT OF vy ...
         AN little_time AN PRODUKT OF 0.5 ...
         AN PRODUKT OF ay AN SQUAR OF ...
         little_time

      vx R SUM OF vx AN PRODUKT OF ax AN ...
         little_time
      vy R SUM OF vy AN PRODUKT OF ay AN ...
         little_time

      tmppos_x'Z i R x
      tmppos_y'Z i R y
      vel_x'Z i R vx
      vel_y'Z i R vy
   IM OUTTA YR loop

   HUGZ

   IM IN YR loop UPPIN YR i TIL BOTH SAEM ...
      i AN 32
      pos_x'Z i R tmppos_x'Z i
      pos_y'Z i R tmppos_y'Z i
   IM OUTTA YR loop

   HUGZ

IM OUTTA YR loop
VISIBLE "O HAI ITZ " ME ", MAH PARTICLZ IZ:"
IM IN YR loop UPPIN YR i TIL BOTH SAEM i AN 32
   VISIBLE pos_x'Z i " " pos_y'Z i
IM OUTTA YR loop

KTHXBYE
```

*E. Compilation and Execution of LOLCODE Programs*

Compilation of LOLCODE occurs via the command line LOLCODE compiler, *lcc*, which translates the code to C and then uses the available C compiler for compilation, linking the OpenSHMEM library to create the final executable program. Some configuration may be needed to specify the location of the OpenSHMEM header and library. Launching the executable may vary by platform. In the case of the Parallella platform, *coprsh* is used to launch the program on multiple cores. For the Cray XC40 supercomputer, *aprun* is used for the same purpose. An example of the relevant commands for compiling and executing a program on 16 parallel processing elements appears below.

```
lcc code.lol –o executable.x
coprsh –np 16 ./executable.x
```

## VII. SOFTWARE AND HARDWARE RESOURCES

The compiler presented in this paper, including additional example codes, will be published under the GPLv3 open-source license. Combined with the $99 Parallella platform we believe this provides a unique resource for teaching parallel and distributed computing concepts in a manner that would be approachable and interesting to students using a programming model that can be executed equally well on a large parallel supercomputing platform.

## VIII. CONCLUSION AND FUTURE WORK

We have implemented language extensions to the internet-meme-based procedural programming language LOLCODE to support parallel and distributed computing within a SPMD PGAS model. The extended programming language remains oddly humorous, in the spirit of the original language, and yet is ironically capable of supporting the implementation of parallel computational algorithms often used in HPC applications. We have presented here the basic elements of the language extensions supporting parallel and distributed computing concepts as well as a complete parallel implementation of a canonical n-body algorithm. In future work we will release the compiler and examples under an open-source license making them available as a teaching resource.